**Title:**

**On-Chip Integrable Planar NbN NanoSQUID with Broad Temperature and Magnetic-Field Operation Range**

**Authors:**

*Itamar Holzman[1,2] and Yachin Ivry[1,2,*]*

**Affiliations:**

1. Department of Materials Science & Engineering, Technion – Israel Institute of Technology, Haifa, 32000, Israel.

2. Solid State Institute, Technion – Israel Institute of Technology, Haifa, 32000, Israel.

Correspondence to: ivry@technion.ac.il

**Abstract:**

Superconducting quantum interference devices (SQUIDs) are used for applications ranging from sensitive magnetometers to low-temperature electronics and quantum computation. We introduce a planar nano SQUID that was made with a single lithographic step out of NbN films as thin as 3 nm on a Si chip. The fabrication process of weak links that are 45 nm in width, and 165 nm in length, which were designed to account for overcoming current crowding are presented. Operation at a temperature range of 20 mK to 5 K as well as at 1 T parallel, and 10 mT perpendicular magnetic fields is demonstrated, while potential operation higher than 8 T has also been shown. The broad range of applicability of a single device as well as its scalability are promising for on-chip integrability that may open new technological possibilities, including in quantum and electro-optical circuiting.

**Text:**

Superconductors are materials that exhibit quantum behavior at the macroscopic scale and hence are attractive in the current race for quantum technologies. Likewise, the lack of resistance in these materials facilitates them for low-power devices, while it also allows them to demonstrate high sensitivity to low-power electromagnetic radiation, which in turn is advantageous for detection technologies. Prominent examples include logic devices,[1,2] gamma and x-ray photon sensors[3–6] and IR single-photon detectors.[7–9] These technologies rely vastly on thin films of superconductors, in which the superconducting properties are weakened. The platform of thin films allows potential integration of superconducting devices with various quantum and non-quantum technologies on the same chip for expanding the advantageous properties from device scale to circuit scale. However, this potential is still unfulfilled, mainly due to the inability to integrate several of the main building blocks, which today involve complex processing procedures. Therefore, there is a strong motivation to introduce such building blocks that are fabricated in a simple manner and allow convenient integration with other superconducting technologies.

Superconducting Quantum Interference Devices (SQUIDs) are widely used in applications, such as magnetic sensing of a broad magnetic-field range and electronic amplification.[10] Being building blocks for solid-state quantum-computation elements,[2,11,12] there is a strong desire to miniaturize the SQUID size for increasing the Qubit density, especially when operated at low temperatures (< 100 mK). Miniaturizing the SQUID size helps also increase its sensitivity to magnetic fields, which is proportional to the SQUID area,[13,14] while for these applications, higher operating temperature is preferable. Moreover, there is a motivation to reduce SQUID size for enabling integrated superconducting and quantum circuits.[15] Nevertheless, common SQUIDs require 3D fabrication of nanoscale tunneling junctions.[10,16,17] Because tunneling junctions reduce significantly the critical current density ($J_c$), the signal-to-noise-ratio in these geometries is often reduced below a reasonable operational limit. Furthermore, this fabrication process is challenging technologically, while the need in a very well-controlled material stack restricts the diversity of materials that can potentially be used, and hence the control over functional properties or the integration in other superconducting technologies.

Nano-fabrication developments allowed for replacement of the commonly used 3D SQUID structure with a planar-SQUID geometry. Here, there is a superconducting ring that is made of a thin film by means of electron lithography, while within the ring, there are two narrow regions that are connected in parallel and serve as the interfering weak links of the SQUID.[18] In addition to the advantageous fabrication process, the geometry of planar SQUIDs is expected to help them sustain very strong magnetic fields along the direction parallel to the device, operate under strong perpendicular fields and exhibit high spin sensitivity, which are all technologically relevant.[19] The sustainability to parallel fields arises thanks to their relatively small cross section, which in turn is obtained because of the thin film thickness and small size of the weak links.[20] That is, given the dependence between the external magnetic flux ($\Phi_{ext}$) and the internal magnetic induction ($B$): $\Phi_{ext} = \int_0^A B \cdot dA'$, for a given critical magnetic flux density, $B_c$, smaller cross-section areas ($A$) leads to sustainability to higher $\Phi_{ext}$. Likewise, the smaller washer area, $a$ (i.e. the area enclosed within the ring) of such geometries allows them to operate well under strong external perpendicular magnetic fields. The reason being is that the breadth of each period in the interfering pattern of the device critical current is proportional to the perpendicular flux, which in turn is proportional to $a$ according to the above integral. Finally, the small washer area gives rise also to high spin sensitivity of the SQUID: $S_n = \frac{2a\Phi_{ns}}{\mu_B \mu_0}$, where $\Phi_{ns}$, $\mu_B$ $\mu_0$ are the system flux noise, Bohr magnetron and vacuum permeability, respectively.[18]

SQUID properties are material depended.[21] To-date, planar SQUIDs have been demonstrated for Nb,[20,22] NbN,[23,24] Pb,[25] YBCO,[26,27] Ti,[28] MgB$_2$,[29,30] and for doped diamond.[31] Amongst these materials, NbN is rather attractive because it has relatively high critical current density and can sustain high magnetic fields, while it is also commonly used for logic, and photon-sensing devices.[6,8,38,39,9,23,32–37] However, NbN has a rather short coherence length ($\xi \approx 6$ nm[40]), challenging the weak-link fabrication, while growing ultra-thin NbN films with controllable properties is also a non-trivial task.[41] Thus far, NbN planar SQUIDs have already demonstrated sustainability to high parallel fields ($\leq 8$ T),[42] sensitivity to perpendicular fields ($\lesssim 3$ mT,[23] note that earlier works discuss sensitivity up to 10 mT, but without a demonstration[24]). Moreover, the relatively high $T_c$ of NbN is attractive for operation at temperatures as high as liquid-helium. Indeed, NbN planar SQUIDs have been reported as operating at such temperatures (3.5 – 5 K).[23,24,42] Nevertheless, as opposed to the advantageous behavior at high temperatures, there have been some concerns regarding the operation of planar SQUIDs at low temperatures. It has already been suggested that NbN planar SQUIDs are sensitive to magnetic fields at Qubit-relevant temperatures (40 mK[24]), but the interference SQUID characteristics have not yet been presented for such conditions. Despite the need in having all of these attractive properties in a single device, to-date, each of these characteristics have been demonstrated for a different system, questioning the ability to have such a single robust device that operates under these various extreme conditions.

Another obvious advantage of the simple fabrication of planar SQUIDs is their potential as integrated devices in superconducting and quantum technologies that are based on thin films. Superconductors usually suffer from miniaturization because that their functional properties are suppressed with thickness reduction.[41] Thanks to the relatively high $T_c$ of bulk NbN, the functional properties of this material are maintained also at ultra-thin films (< 5 nm), where many of the competing material already lose their functionality at technologically relevant temperatures. Hence, NbN is attractive for integrated thin-film technologies, such as superconducting nanowire single-photon detectors (SNSPDs), which detect photons at the communication wavelength with high speed and high efficiency and are based on 4-nm thick films.[7,8] NbN SQUIDs are usually 3D structures that require a very complex fabrication process.[37,43] NbN planar SQUIDs have been reported for ~15-nm films[42] (6-nm thick devices have been discussed, though no interference pattern has been demonstrated[24]), but not for thinner films. Moreover, to increase the device density, smaller footprint (washer area) is needed, while indeed NbN SQUIDs with 750-nm inner radius have already been developed.[42] We should note that some of the other materials[25,26,28,29,31] may exhibit some of the above advantageous properties, but NbN can potentially have them all in one device. Therefore, there is a need for a robust NbN nano SQUID that endures and is sensitive to a broad range of magnetic fields, operates both at low and high temperatures and encompasses thickness and lateral dimensions as well as a fabrication process that are attractive for integration.

Here, using a single-step lithography process, we fabricated on a Si chip, planar 4-nm thick and 500-nm wide NbN planar SQUIDs that operate at a 20 mK – 5 K temperature range and detect reproducibly 100 Gauss perpendicular magnetic fields and up to 1 T parallel fields, while they endure and can potentially operate above 8 T. We tested seven devices out of the twenty devices that we fabricated and they all demonstrated good SQUID behavior. We first sputtered the ultra-thin NbN layer on a Si chip with a thick (338 nm) top layer of amorphous SiO$_2$ that had the role of reducing lattice-mismatch strain in the superconductor. Ellipsometry was used to determine a 4-nm thickness of the NbN layer as demonstrated in Figure SI1. The critical temperature, transition width, residual rate ratio and sheet resistance were $T_c = 9.29$ K, $\Delta T_c = 1.2$ K, $RRR_{@20K} = 0.92$ $[= \frac{R_{@130K}}{R_{@20K}}]$ and $R_s = 157$ $\Omega/\square$ as extracted from Figure SI2, in agreement with the existing literature regarding δ-NbN as was also confirmed with X-ray photoelectron spectroscopy, (Figure SI3). To fabricate the devices, a PMMA layer was spin coated on the sample and a single electron-beam write was done (Raith EBPG 5200, with writing conditions of 100 kV acceleration voltage, 150 pA beam current and 530 $\frac{\mu C}{cm^2}$ overall dose). After developing the resist with MIBK, the pattern was transferred to the sample by means of reactive ion etching (RIE). To overcome the PMMA degradation under the RIE, we avoided accumulated heating in two ways: we cooled the sample holder in the RIE to 1º C and we divided the etching to 20-second steps, allowing the resist to cool down and hence maintain mechanical stability. Figure 1a shows a planar NbN nano SQUID with two parallel weak links of 165 nm length and 45 nm in width and with a 525 x 525 nm$^2$ area. The weak links had a circular geometry to avoid current crowding[44] (see Figure 1b).

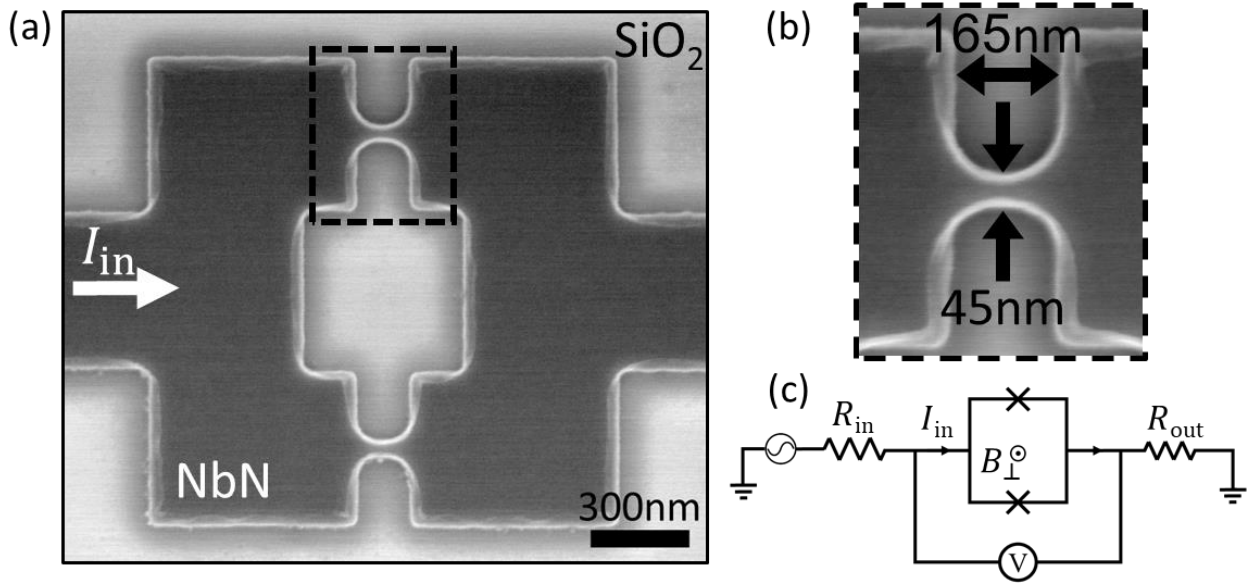

**Figure 1| Planar NbN 4-nm thick nano SQUID.** (**a**) Electron micrograph of the on-chip planar NbN SQUID. (**b**) A closer look at the weak link, demonstrating its current-crowding compensating geometry and dimensions. (**c**) Schematics of the SQUID and the experimental setup circuiting.

To inject current to the device, $I_{in}$, a voltage was applied on a large resistor (~100 kΩ) that was connected in series to the device and the resultant voltage was then measured with a four-probe scheme. Schematics of the electric circuit is given in Figure 1c. Figure 2a shows the interfering magnetic-field – critical-current characteristics between -100 Gauss and 100 Gauss of a perpendicular field ($B_\perp$, limited by the maximal range of our system) even at temperatures as high as 5 K. The interference behavior at 20 mK is given in Figure SI4. Here, the measured critical current of the device was $I_c$ =20.5 µA (all critical current and temperature values are defined as the 90% drop of the normal resistance), while the cycles were of ~6% modulation and ca. 30 Gauss periodicity, which in turn corresponds to an effective washer area of 815 x 815 nm². The device normal resistance $R_n = 1.7$ kΩ was extracted from the resistive region of $V - I$ curve (see Figure 3), so that the device $I_c \cdot R_n$ characteristic, which is an upper limit of the potential magnetometry performance[45] is 34.85 mV. Here, the interference pattern is of a zig-zag form as expected from SQUIDs with long ($l \gg \xi$) weak-link bridges.[46] Measuring the slope of the linear lines of the zigzag pattern allows us to extract the effective inductance, which is the sum of both the kinetic and geometric inductance

$L=8.035\times10^{-10}$ H, while the figure of merit of the device, $\beta_L \equiv \frac{2I_cL}{\Phi_0} = 15.9$, where $\Phi_0$ is the magnetic flux quantum. We should note that devices made of films as thin as 3 nm also demonstrated good interference behavior (Figure SI5).

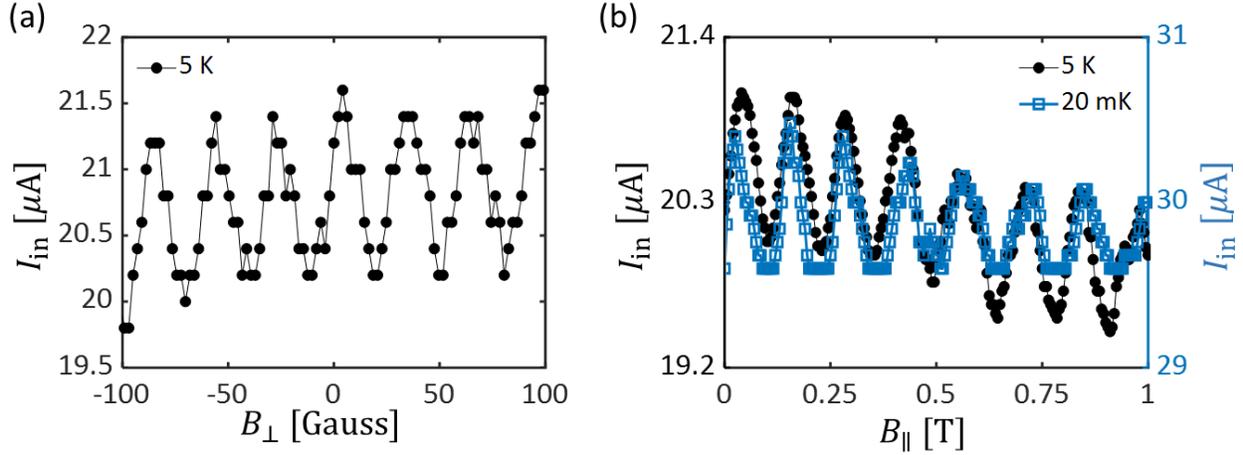

**Figure 2| Planar thin NbN nano SQUID operation at various magnetic fields and temperatures. (a)** Dependence of the switching current on an external magnetic field perpendicular to the sample plane at 5 K (a similar measurement at 20 mK is given in Figure SI4). **(b)** Interference of the same device under high parallel magnetic fields at 5 K (full black circles) and at 20 mK (empty blue squares) showing a similar qualitative behavior, albeit with higher $I_c$ for the lower temperature.

To examine the device applicability range and robustness, we tested its operation under even higher magnetic fields. Because the SQUID is only ca. 4-nm thick, it is expected to be less sensitive to parallel fields $(B_\parallel)$ than it is to perpendicular fields. Figure 2b shows that the device operates reliably as a SQUID up to $B_\parallel=1$ T. The reason that the device was sensitive to parallel fields is most likely because the field was not exactly parallel to the substrate. Given the change in modulation, which has now become 0.133 T, we can calculate the field-substrate offset angle as 1.28°. Yet, the modulation depth was suppressed with respect to the perpendicular-field measurement to a value of 4-5%. To demonstrate the potential of this device, we tested its operation at lower temperatures. Figure 2b shows the device operation under even more extreme conditions of high magnetic field and very low temperatures (20 mK), demonstrating its robustness and broad range of applicability. Here, the switching-current profile followed the higher-temperature measurement (5 K), only the value of the critical current increased from 20.5 µA to 29.4 µA.

Increasing the parallel magnetic field above 1 T introduced significant noise in our system, giving rise to the disappearance of the interference pattern. We believe that the reason being is the instability of our magnetic field (i.e. its current source) rather than the NbN SQUID itself. That is, the large fluctuations in $B_{\parallel}$ affected badly the signal-to-noise ratio so that it was higher than the modulation and hence we could not trace a measurable interference signal. Yet, we believe that under more stable magnetic-field conditions, the device may operate even under much higher magnetic-field values. To support this argument, we measured the dc switching current of the device for different magnetic-field values. Figure 3a shows that even under 8 T (limited by the maximum value of $B_{\parallel}$ in our system), the device $I_c$ was suppressed only by 16.7% (note that a ~1 µA difference was measured between Figure 2 and Figure 3 due to the difference in the less sensitive setup that was used for the $I - V$ characterization in Figure 3). The dependence of $I_c$ on the parallel magnetic field is demonstrated more clearly in Figure 3b (black circles), where we extracted the $I_c$ values from the graph in Figure 3a. These results imply that under a more stable magnetic field and electronics, the device can operate at much higher magnetic fields than we demonstrated in Figure 2b. To further support this claim, we also characterized the dependence of the retrapping current ($I_r$) on magnetic field (blue empty squares in Figure 3b). That is, the value $I_c$ is extracted from the transition from a superconducting state to a normal state along the $I - V$ measurement. Contrariwise, the reverse transition from normal to superconducting occurs at a lower $I_{in}$ value due to a latching effect,[47] which reflects dissipation of the excess heat power that is originated from the current flow in the resistive device. The stability of $I_r$ over a broad range of magnetic fields (Figure 3b) and temperatures (Figure SI6 and SI7) demonstrates the robustness of the device. We should note that the proximity between $I_c$ and $I_r$ for high magnetic fields also indicates on the robustness of the device, while the device retrapping current can be increased significantly, to match the value of $I_c$ and avoiding the hysteretic effect, when a shunt resistor is added in parallel to the weak links.[48,49]

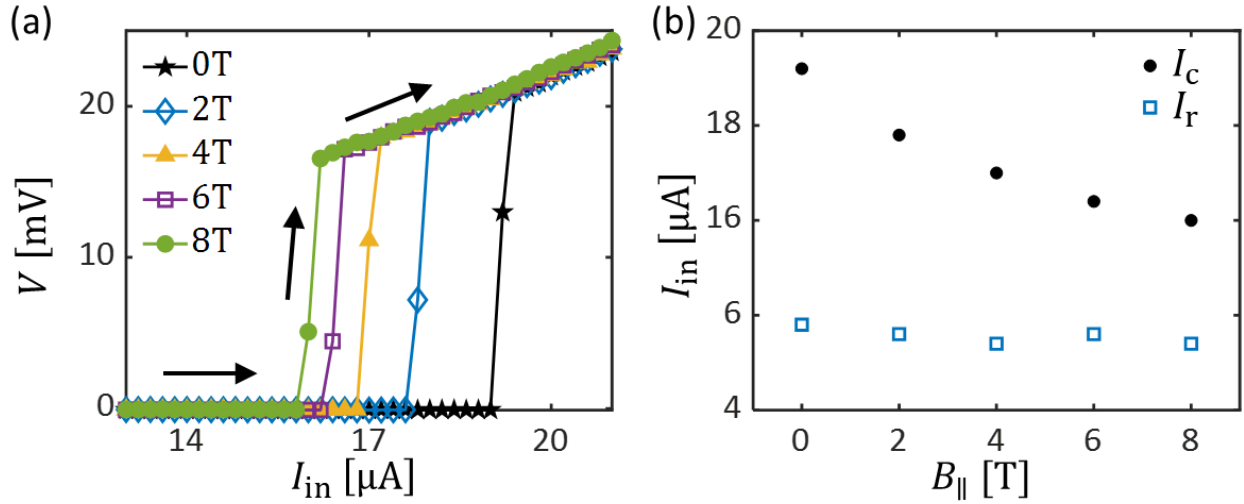

**Figure 3| Current-voltage characteristics of the planar NbN nano SQUID at various magnetic fields.** (**a**) Voltage-current measurements of the device under various parallel magnetic-field values (up to the 8-T limit of our system). (**b**) Weak suppression of $I_c$ (full black circles) and constant retrapping-current ($I_r$) value (empty blue squares) throughout this broad magnetic-field range indicate on the stability of the device and its potential applicability at magnetic field values much higher than those presented in Figures 2 and 3. All measurements were done at 5 K. The complete individual $I-V$ curves that contain both $I_c$ and the retrapping current that were measured as a function of magnet field both at 5 K and at 20 mK are given in Figures SI5 and SI6.

In summary, the device presented here demonstrates that a high-density, on-chip nano SQUIDs that operates under diverse conditions can be made with a simple single lithographic fabrication process from films as thin as 3 nm. The nano SQUID operates at a broad range of perpendicular ($B_\perp$=100 Gauss) and parallel ($B_\parallel$=1 T) magnetic fields, indicating on its applicability as a robust magnetic sensor. The device, which was fabricated on a Si chip operates also at a broad range of temperatures (20 mK to 5 K), demonstrating its applicability for both metrology and quantum-information systems. Finally, the simple fabrication process of the 4-nm thick NbN device is attractive also for integrated devices, including e.g. superconducting nanowire single-photon detectors.

**Supplementary material**

See supplementary material for the NbN layer characterization, detailed fabrication process and complementary measurements.

**Acknowledgments:**

The authors acknowledge financial support from the Zuckerman STEM Leadership Program, and the Israel Science Foundation (ISF) grant # 1602/1. Likewise, we would like to thank the following personnel for technical support and fruitful discussions: Dr. Guy Ankonina (sputtering and ellipsometry); Dr. Adi Goldner and Dr. Roni Winik (device fabrication); Dr. Kamira Weinfeld (XPS); as well as Prof. Hadar Steinberg and Mr. Tom Dvir (low-temperature measurements).

**Supplementary Material:**

The fabrication process began with dc-magnetron sputtering of a 4-nm thick NbN film on top of a Si substrate with a top oxide layer of 338 nm. The reactive sputtering (AJA) of the superconducting $\delta$-NbN phase followed our previous work.[50] The 44-Å film thickness was measured with variable-angle spectroscopic ellipsometry (J. A. Woollam Co Inc. USA, see Figure SI1 for details). After the deposition, PMMA 950A3 electron-beam resist was applied by spin coating, followed by 2-min post baking. A thin layer of 'e-spacer' was used to avoid charging under the electron microscope. Next, the film was patterned with electron-beam lithography (530 $\frac{\mu C}{cm^2}$ dose of 150 pA at 100 kV, Raith EBPG 5200) and the PMMA was developed. Reactive-ion etching (RIE) with $CF_4$ gas transferred the pattern to the NbN layer. To prevent degradation of the PMMA mask, the RIE process was done on a cold stage (1° C). In addition, because the etching process generates heat, we split the process to 3 x 20-sec segments, which allowed the PMMA to cool down. The PMMA was then removed using acetone. Finally, the chip was glued to a holder and wire-bonded with aluminum wires. $T_c$, $\Delta T_c$, and normal-resistance properties were measured from cooling and heating curves (Figure SI2). Material composition and stoichiometry were characterized with x-ray photoelectron spectroscopy (XPS, Figure SI3).

SQUID characterizations were done in BF-LD 250 dilution refrigerator (BlueFors Cryogenics, Finland) with a base temperature of 20 mK. Electric biasing was done with MFLI Lock-in Amplifier (Zurich Instruments, Switzerland). To generate current bias ($I_{in}$) that allows reliable statistical averaging we measured each point at the $I_{in} - B$ graphs with a low-frequency (70-217 Hz) ac voltage that was applied on a 100-kΩ resistor, which was connected in series to the SQUID. Each current value of the $I_{in} - B$ plots were obtained by increasing the amplitude of the ac voltage to that value, while we maintained the sinusoidal ac signal between zero and $|I_{in}|$ by adding dc voltage to the ac signal. The alternating bias allowed us to perform large amount of reproducible measurements, approximately 22 cycles per pixel, as well as to overcome latching, and overcome the lack of shunt resistor in the device. The voltage across the device was amplified (x10) at room temperature before it was measured by the lock-in amplifier. Two superconducting coils and power supplies were used to determine the in-plane ($B_\parallel$) and out-of-plane ($B_\perp$) magnetic fields. $\Delta T_c$ was measured from 90% to 10% of the device normal resistance. *RRR* was defined as the ratio between the device resistance at 130 K and 20 K. $R_s$ was measured for the deposited layer before patterning with 4 probes (Van der Pauw technique).

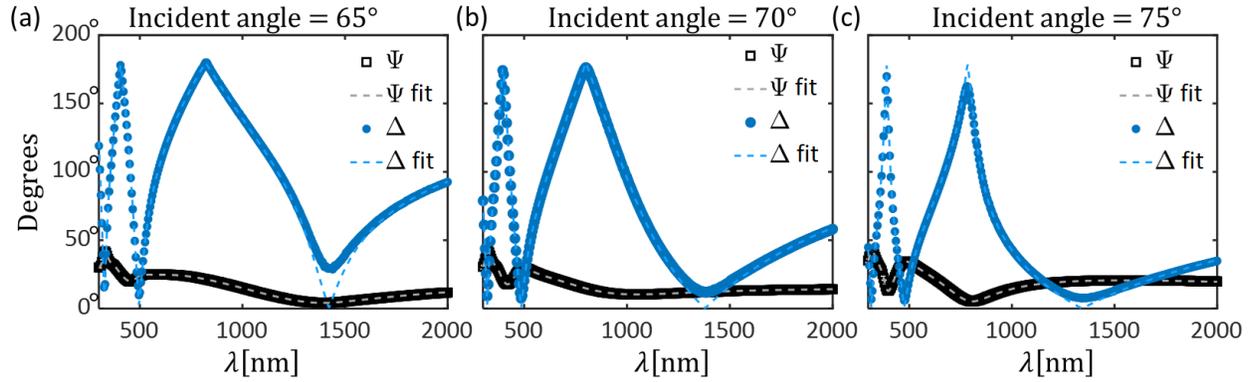

**Figure SI1| VASE measurements of the complex reflectance ratio of an NbN chip at different incident angles**. Ellipsometry parameters ($\Psi$ and $\Delta$) of the complex reflectance ratio between the primary and secondary beam: $\rho = \tan(\Psi)\, e^{i\Delta}$ as a function of the wavelength for (**a**) 65°, (**b**) 70° and (**c**) 75° incident angles between the beam and the sample. Dashed lines are the model fit. Here, we used a five-oscillator fit

(one Drude, one Tauc-Lorentz, one Gauss-Lorentz and two Gaussians). Using this model to fit the data we extracted the NbN layer thickness of 4.4 Å.

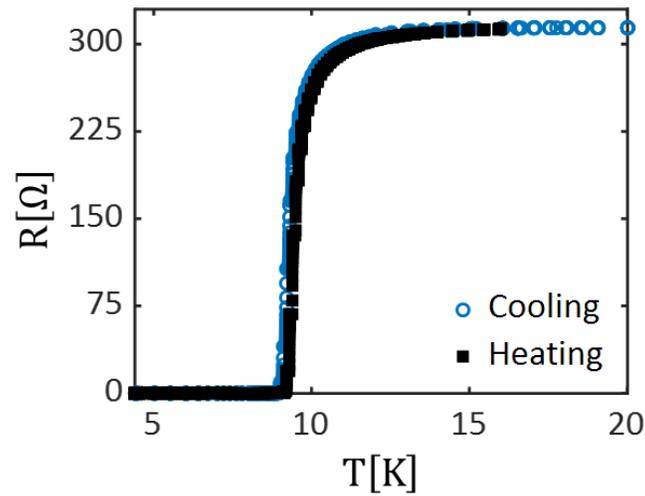

**Figure SI2| Cooling and heating curves of the NbN chip**. Resistance of the NbN film was measured in a 4-probe configuration as function of temperature while both cooling and heating the sample. The sample resistance at 20 K was 314 Ω and the $T_c = 9.29$ K was determined as the temperature at which the resistance dropped to 10% of $R_{@20K}$. The transition width $\Delta T_c = 1.2$ K was measured as the temperature difference

between the point at which $R = 0.9 \cdot R_{@20K}$ and the critical temperature. The ratio between the residual resistance ratio between the resistance at 130 K and $R_{@20K}$ and was found to be close to unity: $\rho = 0.92$, while the reproducibility of the results between the cooling and heating implies on the quality of the superconductor.

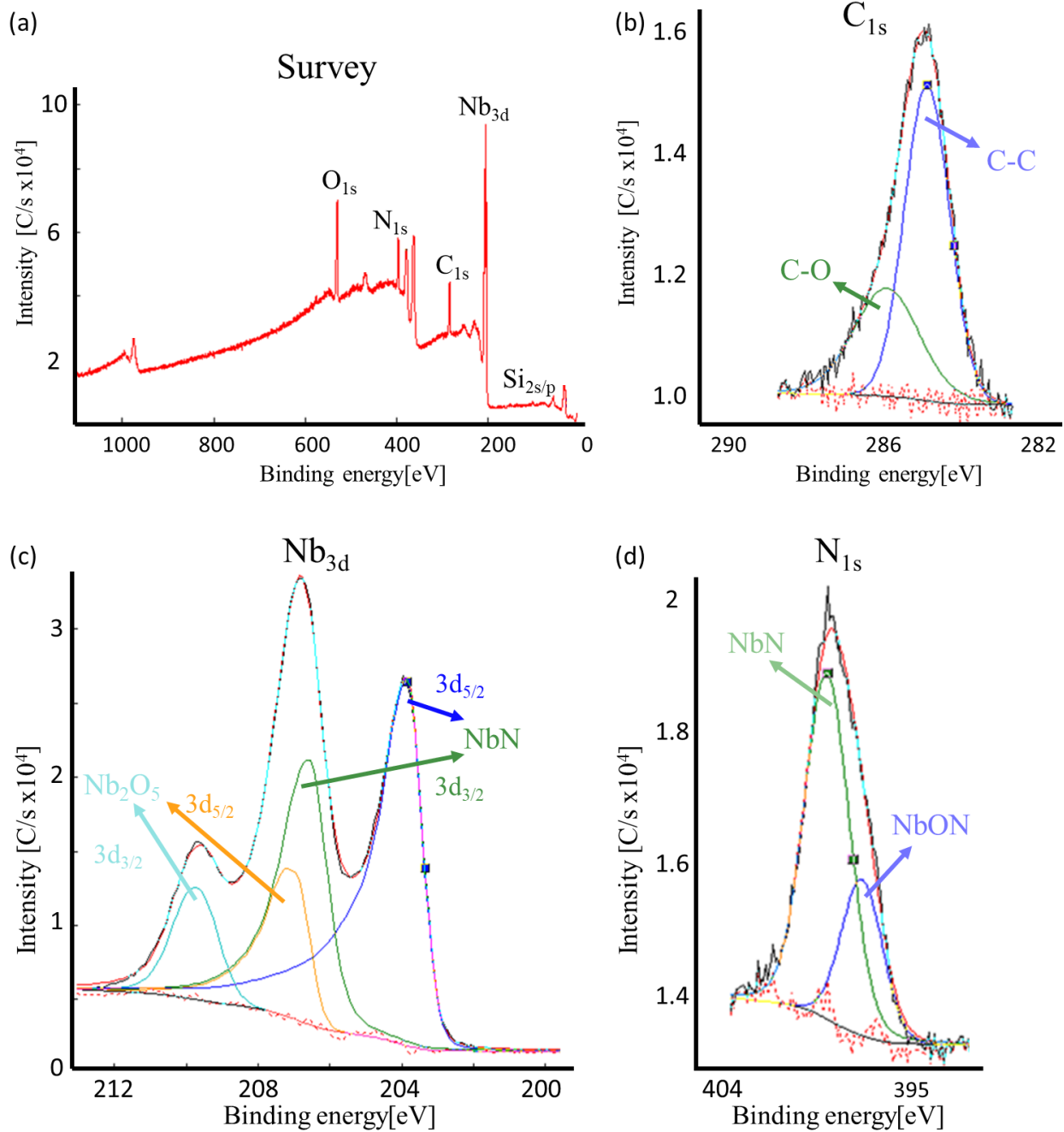

**Figure SI3| XPS analysis of the film composition and stoichiometry.** (**a**) Element survey of the as-deposited film showing the presence of Nb (23.3 at%) and N (17.2 at%) along with presence of O (29.3 at%) and C (29.0 at%) and slight amount Si (1.2 at%) from the substrate (note that the sample was transfer from the growth system to the XPS in air). (**b**) High-resolution measurement of the carbon peak (black dots) of the as-deposited sample indicates on the existence of organic contamination on the surface and was used for calibrating the energy of the XPS measurements. C-C bonds were measured at 284.8 eV (purple line is the fit) and C-O bonds at 285.9 eV (green line is the fit), red dots are the background (using the Shirley algorithm[51]) and light-blue line is the overall fit $\chi^2 = 0.994$.[52–54] (**c**) High-resolution XPS measurement around the $Nb_{3d}$ and (**d**) $N_{1s}$ bonding energy. The peak positions were determined by using a Voigt fitting, which took account the spin-orbit split. Metallic NbN bond ($d_{5/2}$) was found at 203.85 eV,[55,56] while a peak ($d_{5/2}$) of a $Nb_2O_5$ oxide layer was found at 206.97 eV[57,58] ($\chi^2 = 5.084$ was found as the fit error). The $N_{1s}$ spectrum consists of mainly metallic NbN bond at 397.2 eV[55,56] and an NbON peak at 396.3eV[59] ($\chi^2 = 1.991$). Black dotted line is the overall sum of the measured data, red dotted line is the background (found with the Shirley algorithm), solid red line is the overall fit of the combined data while the colored lines correspond to different phases are fitted peaks as designated in the figure (Green – NbN (c) NbN $3d_{3/2}$ (d), purple – NbON (c) NbN $3d_{5/2}$ (d), light blue – $Nb_2O_5$ $3d_{3/2}$, orange – $Nb_2O_5$ $3d_{5/2}$, note that the other colors are an artefact that was produced by the software).

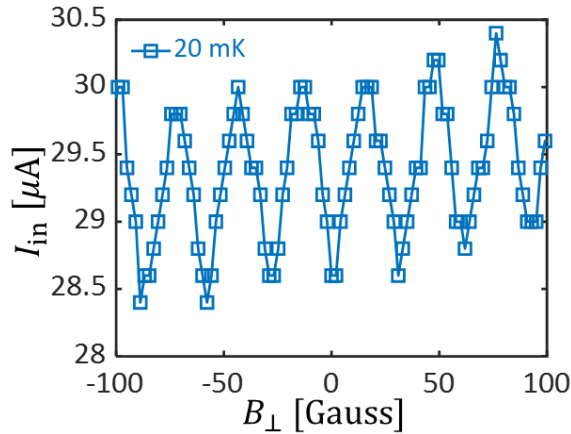

**Figure SI4| Planar thin NbN nano SQUID operation at 20 mK.** The switching current pattern is shifted in respect to the measurement at 5 K (Figure 2) by half a cycle along the magnetic field axis and by 8 μA along the current bias axis.

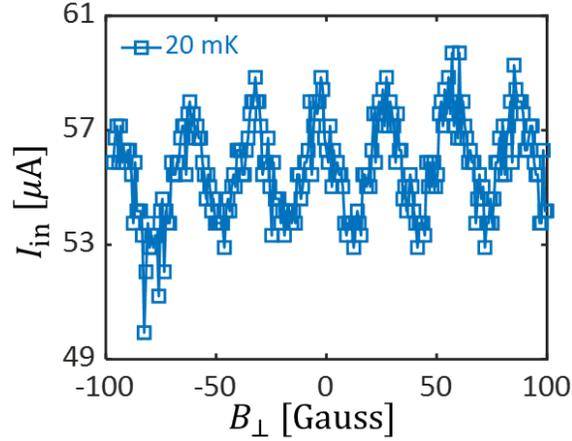

**Figure SI5| 3-nm thick planar NbN nano SQUID.** Switching current pattern of a 3 nm thick NbN SQUID operating at 20 mK. The switching current of the device is $I_c = 56$ µA, while ~9% modulation is observed along c.a 29 Gauss period.

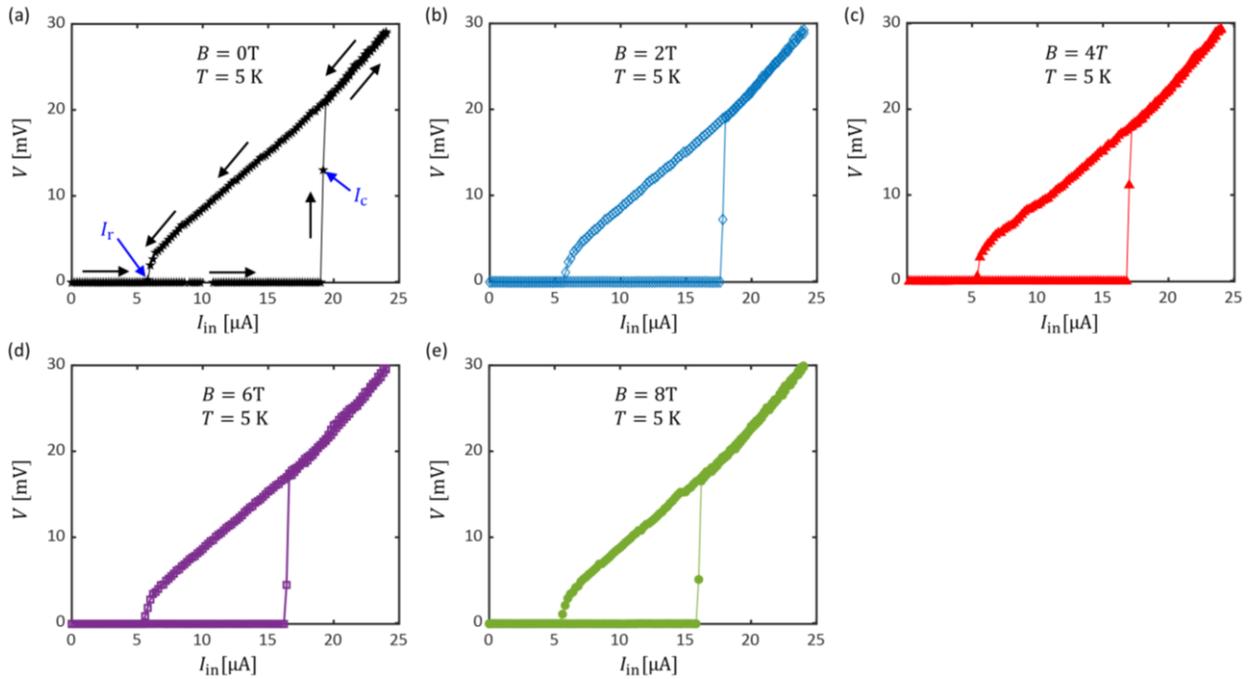

**Figure SI6| Voltage-current curves of the SQUID at 5 K for varying parallel magnetic-field values.** Measured voltage across the device that corresponds to dc bias current cycling from 0 µA to 25 µA and back to 0 µA at (**a**) 0 T, (**b**) 2 T, (**c**) 4 T, (**d**) 6 T and (**e**) 8 T parallel magnetic fields. The extracted critical currents and retrapping currents are presented in Figure3b.

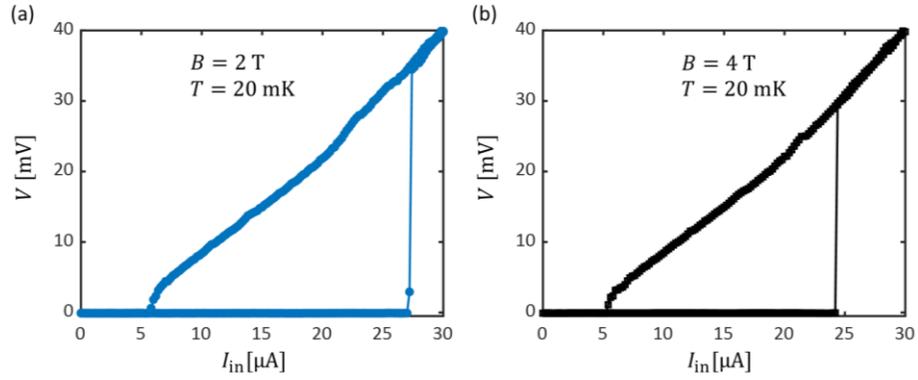

**Figure SI7| Voltage-current curves of the device operating at 20 mK for different parallel magnetic fields.** Measured voltage across the device that corresponds to dc bias current cycling from 0 µA to 25 µA and back to 0 µA at (**a**) 2 T, (**b**) 4 T parallel magnetic fields. These characteristics are similar to the measurements at 5 K, only with an expected increase in $I_c$ ($I_r$ has remained unchanged).